\documentclass[prl,superscriptaddress,showpacs,twocolumn]{revtex4}

\usepackage{graphicx,amsmath,amsfonts}
\usepackage{verbatim}

\newcommand{\figwidth}{0.8\columnwidth}
\newcommand{\uhat}{\hat{u}}
\renewcommand{\Im}{{\rm Im}}
\renewcommand{\Re}{{\rm Re}}

\newcommand{\fref}[1]{Fig.~\ref{#1}}

\begin{document}

\title{Dynamic Compressibility and aging in Wigner crystals and
quantum glasses}

\author{Leticia F. Cugliandolo}
\affiliation{Laboratoire de Physique Th\'eorique et Hautes Energies,
4 Place Jussieu, 75252 Paris Cedex 05, France}
\affiliation{LPTENS CNRS UMR 8549 24, Rue Lhomond 75231 Paris
Cedex 05, France}
\author{Thierry Giamarchi}
\affiliation{University of Geneva, DPMC, 24 Quai Ernest Ansermet,
CH-1211 Geneva 4, Switzerland}
\author{Pierre Le Doussal}
\affiliation{LPTENS CNRS UMR 8549 24, Rue Lhomond 75231 Paris
Cedex 05, France}

\date{\today}

\begin{abstract}
We study the non-equilibrium linear response of quantum elastic systems
pinned by quenched disorder with
Schwinger-Keldysh real-time techniques complemented by a mean-field
variational approach. We find (i) a quasi-equilibrium regime
in which the analytic continuation from the imaginary-time replica results
holds provided the marginality condition is enforced; (ii) an aging
regime. The conductivity and compressibility are computed. The latter is
found to cross over from its dynamic to static value on a scale set by the
waiting time after a quench, an effect which can be probed in experiments
in {\it e.g.} Wigner glasses.
\end{abstract}
\maketitle

The dynamics and transport properties of glasses
 are the object of
current theoretical and experimental
interest~\cite{vincent_sitges_review,cugliandolo_dynamics_leshouches,giamarchi_book_young}.
The slow approach to the static limit and the accompanying aging
phenomena~\cite{vincent_sitges_review} observed experimentally in
classical glasses is captured by a variety of
models~\cite{cugliandolo_dynamics_leshouches}.  A natural explanation
for the slow dynamics is the existence of a special organization of an
exponentially large
 number of metastable states.

Less is known when
glasses evolve at very low temperatures and
 quantum fluctuations
become important as occurs in
Wigner~\cite{andrei_wigner_2d,giamarchi_wigner_review} and
Coulomb~\cite{ovadyahu_coulomb_glass_review,grempel_coulomb,vlad_coulomb_glass,muller_coulomb_glass}
glasses, as well as in spin~\cite{rosenbaum_spin_glass} and other
systems~\cite{osheroff_aging_spinglass} at low
 temperature. Aging in
the transport properties of Coulomb glasses was reported
recently~\cite{ovadyahu_coulomb_glass_review}. Theoretically,
electronic glasses have been studied with numerical simulations in a
classical limit~\cite{grempel_coulomb} or with the imaginary-time
methods combined with the replica
trick~\cite{vlad_coulomb_glass,muller_coulomb_glass} applied to the
isolated model. Strictly, the
latter allows one to describe the statics and, {\it via} an analytical
continuation to real time, the equilibrium dynamics
with infinitesimal dissipative coupling to a bath.
However, it is of a somehow broader use since it has also given access to
dynamical quantities such as the finite frequency
conductivity~\cite{giamarchi_columnar_variat,chitra_etal,fogler_pinning_wigner},
and instanton calculations allowed to relate, in the presence of a
bath, the
 imaginary time solution to the ultra-slow dynamics
(quantum creep)
 \cite{nattermann_temperature_luttinger}. Still, a
rigorous
 treatment of a full non equilibrium relaxation requires
special techniques
 devised to deal directly with the real-time
dynamics of dissipative quantum
systems~\cite{cugliandolo_dynamics_quantum,cugliandolo_dynamics_quantum_aging}.
Even in the infinitesimal coupling limit these are needed
 to ascertain
the validity of analytic continuations to real-time, especially for
glassy systems.

The compressibility of electronic glasses has been measured recently
\cite{eisenstein_hall_compressibility_long,dultz_compressibility_wigner,ilani_compressibility_2DEG}.
Being of thermodynamic nature, in
 equilibrium it
should only depend on the statics of the problem.
 Conversely, its time
dependence reflects an out of equilibrium
 relaxation.  Within an
imaginary-time (Matsubara) variational
 calculation
\cite{giamarchi_columnar_variat,chitra_etal,chitra_wigner_zerob} the
static, zero frequency, compressibility of a disordered Wigner
 crystal was
found to be non-zero and identical to the one of the
 pure system. In
contrast, the analytic continuation yielded a
 vanishing result even in the
small real frequency limit. This hinted
 to the fact that aging effects could
be present
 \cite{giamarchi_wigner_review,chitra_wigner_zerob} though a
firm
 conclusion was clearly beyond the imaginary-time calculation that
assumes equilibrium at the outset.

 In this Letter we study the out of
equilibrium relaxation of
 a disordered Wigner crystal with the
Schwinger-Keldysh (SK) technique
\cite{keldysh1965,kamenev_keldysh_review}. We find two two-time regimes; one in which
the equilibrium result is
recovered and another one in which
 the compressibility
is reduced and dominated by aging effects. Our
 calculation is performed
within a mean-field-like variational approximation.
 We discuss its limits of
validity as well as the relevance of our results
 for experimental systems.

 We model a disordered quantum crystal as
\begin{multline} \label{eq:depart}
 H = \int_x \left[\frac{\hbar^2 \Pi^2(x)}{2m \rho_0} + \frac{c}2 (\nabla u(x))^2
 \right] \\
 - \int_x \; U(x) \rho_0 \cos[Q (x - u(x))]\; ,
\end{multline}
where $m$ is the mass of the particles, $\rho_0$ is the average
density, $Q \equiv 2\pi/a$ with $a$ the inter-particle spacing and
$\int_x\equiv \int d^dx$. $\Pi$ and $u$ are conjugate operators
$[u(x),\Pi(x')] = i \hbar\delta(x-x')$. $U(x)$ is a random potential
with Gaussian statistics, $\overline{U(x)}=0$ and
$\overline{U(x)U(x')} = \Delta(x-x')$ with $\Delta(z)$ a function
with finite range $r_f$. This model describes a large class of
systems including charge density waves
\cite{gruner_book_cdw,fukuyama_pinning} (with a phase $\phi = 2\pi
u/a$), Wigner crystals (upon generalization to a two component
vector $\vec u$) \cite{chitra_etal,giamarchi_wigner_review}, and
Luttinger liquids in $d=1$ \cite{giamarchi_book_1d}.

The compressibility $\kappa$ is defined as the change in density in
response to a change in chemical potential $\mu$. In linear
response, it is given by the $q\to 0$ limit of the density-density
correlator. For (\ref{eq:depart}) the long wavelength part of the
density is $\rho(x) -\rho_0 \sim -\rho_0 \nabla u(x)$, and the
equilibrium compressibility is
$\kappa = \lim_{q\to 0} \kappa(q,\omega_n=0)$ with
$\kappa(q,\omega_n) =  \rho_0^2 q^2 G_c(q,\omega_n)$, $\omega_n$
the Matsubara frequencies, and
$G_c(q,\omega_n) = \overline{\langle u^*_{q,\omega_n}
u_{q,\omega_n} \rangle - \langle u^*_{q,\omega_n} \rangle \langle
u_{q,\omega_n} \rangle}$, where $\langle\ldots\rangle$ and
$\overline{\cdots}$ denote
thermal  and disorder average, respectively.
Within the replica variational approach
\cite{giamarchi_columnar_variat}
\begin{equation} \label{eq:disorcomp}
 \kappa(q,\omega_n) = \frac{\rho_0^2 q^2}{\rho_m \omega_n^2 + c q^2 +
 \Sigma_1(1-\delta_{n,0}) + I(\omega_n)}
\end{equation}
where $\rho_m = m \rho_0$, $\Sigma_1 \sim \rho_m \omega_p^2$ is a
constant depending on disorder, $\omega_p$ is the pinning frequency
\cite{fukuyama_pinning,giamarchi_columnar_variat} and $I(0)=0$.
$\kappa$ is \emph{independent of disorder} and simply given by
\begin{equation} \label{eq:compstat}
 \kappa = \rho_0^2/c \; .
\end{equation}
Alternatively, the real-time compressibility, {\it i.e.} the
response to a time-dependent chemical potential, is given by the
retarded linear response. Naively, this can be obtained
from (\ref{eq:disorcomp}) by the
standard analytical continuation
$i\omega_n \to \omega + i \delta$
that leads to~\cite{giamarchi_columnar_variat}
\begin{equation}
 \kappa(q,\omega) =
 \rho_0^2 q^2 \; [-\rho_m \omega^2 + c q^2 + \Sigma_1 +
 \tilde{I}(\omega)]^{-1} \; ,
\end{equation}
where $\tilde{I}(\omega\to 0) \to 0$. Note that even performing the
limit $\omega \to 0$ \emph{first} while keeping $q$ fixed one finds
\begin{equation}
\label{eq:compdyn}
 \kappa = \lim_{q \to 0} \lim_{\omega\to 0} \kappa(q,\omega) = 0
\; ,
\end{equation}
in disagreement with the result in (\ref{eq:compstat}).

The simplest example where such a difference arises is an isolated
two level system ({\it e.g.} a spin). The static susceptibility in
response to an external magnetic field is $\chi = 1/T$ where $T$ is
the temperature ($k_B=1$). However, the response to a time dependent
field is always zero, leading to $\chi(\omega\to 0) = 0$ at variance
with the static result. The difference is due to the existence of
two degenerate ground states in the unperturbed system. In the
static calculation one sums over both. The perturbation instead is
unable to induce transitions leading to zero response. In a glass
there may be no exact degeneracy between the multitude of metastable
states. Still, if they are close enough in energy and their coupling
is sufficiently weak one expects a similar phenomenon.

To analyze this issue, we study the real-time dynamics of model
(\ref{eq:depart}) using the SK technique. Let us sketch the main
steps of the calculation. The observables are computed using a path
integral on two
 fields, $u_\pm$, that live on the two sides of a
closed time contour.
 The measure is given by $e^{-S_K}$ where $S_K$
is the
 (dimensionless) SK action $S_K = \frac{i}{\hbar}[S(u_+) -
S(u_-)]$
 and $S(u)$ is the standard action for a system described
by
 (\ref{eq:depart}). To take into account non zero temperature
and
 dissipation we couple the system to a thermal bath
 of
independent harmonic
oscillators~\cite{cugliandolo_dynamics_quantum}. By integrating them out
we induce a coupling between the fields $u_\pm$. The average over
disorder is easily done without
 replicas and has a similar
(coupling) effect. In terms of the fields
 $u \equiv \frac12(u_+ +
u_-)$ and $\uhat \equiv \frac{1}\hbar(u_+ -
 u_-)$ the resulting
action, $S_{av} = S_0 + S_d$, depends on ${\sf R}(z) = \rho_0^2
\Delta_Q \cos(Q z)$ with $\Delta_Q$ the Fourier transform of the
disorder correlation at wavevector $Q$, and reads
\begin{widetext}
\begin{equation} \label{eq:keldaction}
\begin{split}
 S_0 &= \frac12 \int_{q,\omega}
 \left( \begin{array}{cc} u_{q,\omega}^* & i \uhat_{q,\omega}^*
 \end{array}\right)
 \left(\begin{array}{cc}
 0 & -\rho_m\omega^2 + c q^2 + i\eta \omega
\\
 -\rho_m\omega^2 + c q^2 - i\eta\omega &
 -\eta\hbar\omega\coth(\frac{\hbar\omega}{2T})
 \end{array}\right)
 \left(\begin{array}{c} u_{q,\omega} \\ i\uhat_{q,\omega}
 \end{array}\right)
\; ,
\\
 S_d &=   \frac{1}{2 \hbar^2}  \int_{xtt'} \sum_{\sigma=\pm
 1,\sigma'=\pm 1} \sigma \sigma' \; {\sf R}[u_{xt} - u_{xt'} +
 \frac{\hbar}{2} (\sigma \uhat_{xt} - \sigma' \uhat_{xt'})]
\; ,
\end{split}
\end{equation}
\end{widetext}
with $\int_{q\omega}\equiv \int \frac{d^dq}{(2\pi)^d} \int
\frac{d\omega}{2\pi}$. The terms proportional to $\eta$ arise from
the coupling to the bath and represent Ohmic dissipation. For $\eta$
infinitesimal one recovers the intrinsic dynamics of the
quantum system. For the pure system (${\sf R}=0$) this action yields
the equilibrium response obtained by the analytic continuation of
the Matsubara representation. $S_d$ can be rewritten as $S_d =
\frac{2 \rho_0^2 \Delta_Q}{\hbar^2} \int_{xtt'} \sin(Q
\hbar\uhat_{xt}/2)\sin(Q \hbar\uhat_{xt'}/2)\cos[Q(u_{xt} -
u_{xt'})]$; it is clear that the quantum action crosses over to
the dynamic Martin-Siggia-Rose action \cite{chauve_creep_long} when
$\hbar\to 0$.

The disorder-averaged correlation and linear response are defined as
$C_{x-x'}(t,t') = \frac{1}{2}
 \langle u_{xt}^+ u_{x't'}^- + u_{xt}^-  u_{x't'}^+ \rangle = \langle
 u_{xt} u_{x't'} \rangle$ and $R_{x-x'}(t,t') = \frac{i}{\hbar }
 \langle u_{xt}^+ (u_{x't'}^+ -  u_{x't'}^-)\rangle =
 \langle u_{xt} i \hat{u}_{x't'} \rangle$, with $R_{x-x'}(t,t')$ for $t\leq t'$ and $\langle \hat{u}_{xt}
\hat{u}_{x't'} \rangle$ vanishing because of causality. The brackets
represent here an average with the weight $e^{-S_{av}}$. In the absence
of disorder, $C$ and $R$ are stationary, {\it i.e.} depend only on
$t-t'$, their Fourier expressions are
\begin{equation} \label{eq:freecor}
\begin{split}
 C^0_{q}(\omega) &= \frac{\eta\hbar\omega\coth(\frac{\hbar\omega}{2T})}
 {(\rho_m \omega^2 - c q^2)^2 + \eta^2\omega^2}
\; ,
\\
 R^0_{q}(\omega) &=
[- \rho_m \omega^2 + c q^2 -i \eta \omega]^{-1} \; ,
\end{split}
\end{equation}
and obey the fluctuation-dissipation theorem (FDT)
\begin{equation}
 R^0_{q}(\omega) = - \int_{\omega'}
 \frac{2\tanh\left(\frac{\beta \hbar \omega'}{2}\right)
}{\hbar \, (\omega - \omega' + i \epsilon)}
 \; C^0_{q}(\omega')
\; .
\label{eq:FDT}
\end{equation}
In particular, $\hbar \Im R^0_{q}(\omega) = \tanh(\beta
\hbar \omega/2) C^0_{q}(\omega')$. Since the long wavelength part of
the density and the current are respectively $\rho(x)-\rho_0 =
-\rho_0 \nabla u(x,t)$ and $j(x) = \rho_0
\partial_t u(x,t)$,  in linear response
the compressibility and conductivity are given by
\begin{equation}
 \kappa(q,\omega) = q^2 \rho_0^2 \, R_{q}(\omega)
\;, \;\;\;
 \sigma(\omega) = -i \omega \rho_0^2 \, R_{q=0}(\omega) \; .
\end{equation}
Inserting the free values in (\ref{eq:freecor}) one recovers
(\ref{eq:compstat}) for the static compressibility and
$\sigma(\omega) = \rho_0^2/(\eta - i \omega \rho_m)$ for the static
conductivity which, in the limit $\eta\to 0$, reproduces the Drude
form, $\Re \sigma(\omega) = \frac{\rho_0 \pi}m\delta(\omega)$.

In presence of disorder, one derives self
consistent mean-field equations for $C_{q}(t,t')$ and $R_{q}(t,t')$
where $t=0$ is the time when the system is
quenched into the disordered state and set in contact with the bath.
We show only the equation for
the response (see
\cite{cugliandolo_compressibility_quantum_long} for details)
\begin{multline} \label{eq:responseq}
 (\rho_m \partial_t^2 + \eta \partial_t + c q^2)  R_q(t,t') =
 \delta(t-t') \\
 - \int_0^t ds \; \Sigma(t,s) \left[R_q(t,t') - R_q(s,t') \right] \;,
\end{multline}
with the self-energy $\Sigma(t,s) = -\frac4\hbar \Im V'[\tilde
B(t,s) + i \hbar \tilde R(t,s)]$ and $V(z) = -\rho_0^2 \Delta_Q
e^{-\frac12 Q^2 z}$. The tilde denotes an integration over $q$ [{\it
e.g.} $\tilde{R}(t,s) = \int_q R_q(t,s)$] and
$B_q(t,t')=C_q(t,t)+C_q(t',t')-2 C_q(t,t')$.

Analysis of this equation in the long time limit $t,t' \to \infty$
shows that, as in the classical case
\cite{cugliandolo_ledoussal_manifold_global} and the quantum
dissipative $p$-spin
model~\cite{cugliandolo_dynamics_quantum,cugliandolo_dynamics_quantum_aging},
the model exhibits two two-time regimes. First, for fixed $t-t'$
the two-time functions are stationary, $R_q(t,t') \to r_q(t-t')$ and
$B_q(t,t') \to b_q(t-t')$, and the FDT (\ref{eq:FDT}) holds. For
more separated times, there is an asymptotic aging solution,
$R_q(t,t')\to R^A_q(t,t')$, as discussed below. The equation for
$r_q(t-t')$ is obtained from (\ref{eq:responseq}) by a careful
separation of time scales along the lines of
\cite{cugliandolo_pspin}. Its solution reads
\begin{equation} \label{solufdt}
 r_q(\omega) =
[c q^2 + M - i \eta \omega - \rho_m \omega^2 - \Sigma(\omega)]^{-1}
\; ,
\end{equation}
with the self-energy $\Sigma(\omega) =
\int_0^\infty d\tau (e^{-i\omega\tau}-1)
\Sigma(\tau)$ and
\begin{equation}
 \Sigma(\tau) =  \frac{2}{\hbar} \sum_{\sigma=\pm 1} i \sigma \,
 V'[ \tilde{b}(\tau) + i \hbar \sigma \tilde{r}(\tau)] \; .
\end{equation}
The constant $M= \lim_{t \to \infty} \int_0^t  ds \Sigma(t,s) -
\int_0^{+\infty} d \tau' \Sigma(\tau')$ the so called anomaly,
arises from the contribution of the aging time-scales to the FDT
regime. Using FDT,
\begin{equation}
 \tilde b(t) = \int_\omega 2 (1 - e^{- i \omega t})
 \hbar \, [1 + 2 f_B(\omega)] \, \Im \tilde r(\omega) \; ,
\end{equation}
where $f_B(\omega)=1/(e^{\beta \hbar \omega}-1)$. The last self-consistency condition follows
from matching the FDT regime with the aging one. Taking $t' \to t^-$
in (\ref{eq:responseq}) yields $(c q^2 + M) R_q^A(t,t^-) =
\Sigma^A(t,t^-) r_q(\omega=0)$, and the existence of a non vanishing
aging solution with $R_q^A(t,t') \to 0$ as $t' \to 0$ requires the
`marginality condition'
\begin{equation} \label{marginality}
 1 = - 4 V''(b_\infty) \int_q
 (c q^2 +  M)^{-2} \; ,
\end{equation}
where $b_\infty = \lim_{t \to \infty} \tilde b(t)$. One can
explicitly check \cite{cugliandolo_compressibility_quantum_long}
that the solution in the FDT regime coincides with the analytical
continuation of the saddle-point solution of the replica variational
approach to the Matsubara action performed in appendix D of
\cite{giamarchi_columnar_variat}. More precisely, $G_c(q,\omega_n)$
identifies (after analytic continuation) with $r_q(\omega)$.
Similarly, $I(i\omega_n\to \omega + i \delta) \to \tilde
I(\omega)=-\Sigma(\omega)$, $\Sigma_1 \to M$, $B\to b_\infty$ where
$B$ is defined in (31) in \cite{giamarchi_columnar_variat}.
Importantly, this correspondence holds only if the replica symmetry
breaking scheme is the one using the marginality condition
(\ref{marginality}). This choice, advocated in
\cite{giamarchi_columnar_variat} for being the only one leading to a
gapless conductivity, was often used since and is hereby fully
justified within the SK formalism. Note that in $d=1$ and for
$\eta=0$ model~(\ref{eq:depart}) has an additional stable 1 step
replica symmetry breaking solution that is then discarded in the
Matsubara treatment. The SK formalism also allows one to obtain the
response for finite $\eta$. The low frequency behavior of the
conductivity changes from $\Re\sigma(\omega) \sim \rho_0\omega^2/(m
\omega_p^3)$ to $\Re \sigma(\omega) \sim \sqrt{\eta}
\omega^{3/2}/\omega_p^3$ \cite{foot2}.

\begin{figure}
 \centerline{\includegraphics[width=\figwidth]{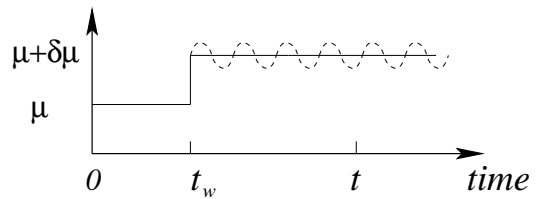}}
 \caption{Sketch of the variation of the chemical potential;
 dc (solid line) and ac (dashed line), for a system quenched
 at time $t=0$. The density is measured at
 time $t$.
 \label{fig:square}}
\end{figure}

We now turn to the compressibility, {\it i.e.} the
(linear) response of the density
to a small change in chemical potential of
amplitude $\delta\mu$ applied between times $t_w$ and $t$ (see
\fref{fig:square}).
One finds
\begin{equation} \label{eq:compag}
 \kappa(q;t,t_w) \equiv
 \left. \frac{\delta\rho(t)}{\delta\mu} \right|_{\mu=0}
 =
 \frac{\rho_0^2 q^2}{c q^2 + M(t,t_w)}
\; .
\end{equation}
As $t-t_w$ increases one distinguishes
two regimes. For $\omega_p^{-1} \ll t-t_w \ll t_w$ the response is
dominated by the `FDT regime' and it is stationary; moreover
$M(t,t_w) = M$ and the compressibility vanishes
[$\kappa(q;t,t_w)\to 0 $ when $q\to 0$]
recovering~(\ref{eq:compdyn}). If, instead, $t-t_w \sim t_w$, the
response is dominated by the aging regime and $M(t,t_w) = M
-4\int_{\tilde{B}^A(t,t_w)}^{\tilde{b}_\infty} V''(z) X[z] dz$,
where $X(z)$ is the FDT violation
ratio~\cite{cugliandolo_dynamics_quantum,cugliandolo_ledoussal_manifold_global,cugliandolo_pspin}
as a function of $z=\tilde
B$. For $\eta$ infinitesimal, $X(z)$ coincides
with $X \to u$, $z \to B(u)$ as found in the replica solution
with (\ref{marginality})~\cite{giamarchi_columnar_variat}.
$\tilde B$ has an aging form
and its detailed scaling depends on the
model. In $d=1,2$ one finds $M(t,t_w) \simeq M F(h(t)/h(t_w))$, where $F$ and
$h$ are scaling functions.
When times are very separated $t-t_w \gg t_w$, $M(t,t_w)$ tends to
zero and, from (\ref{eq:compag}), the compressibility is the constant
(\ref{eq:compstat}). At intermediate  time
scales a mass is always present and the compressibility depends on
the wavevector $q$. For a finite size system of size $L$ one can estimate
that the compressibility crosses over from being essentially
zero to the thermodynamic one when
\begin{equation}
(L_c/L)^2 > M(t,t_w)/M
\; ,
\end{equation}
[$L_c=\sqrt{c/\rho_m} \omega_p$
is the Larkin pinning
 length].
The characteristic timescale to realize such
evolution is
 the waiting time $t_w$. This potentially provides a
direct
 experimental way to check for aging in these
systems. Alternatively,
 one can apply an ac perturbation as
typically done in experiments
 (see Fig.~\ref{fig:square}, and
\cite{vincent_sitges_review} for a
 similar discussion on the ac
magnetic susceptibility of classical
 spin-glasses) and obtain
different results by tuning the
 period, $\tau$, of the
perturbation. Short $\tau$'s select the FDT
 contribution and thus a
vanishing compressibility while very long
 $\tau$'s do not erase the
contribution from the aging regime
 yielding a constant
compressibility as in (\ref{eq:compstat}). The
 crossover in $\tau$
is typically of the order of $t_w$.

 The above results are
consistent with the picture that upon a change
 of chemical potential
the system is first trapped in
 quasi-equilibrium in a metastable
state. Since these states are
 pinned the charge cannot fluctuate and
the compressibility is
 essentially zero. On the other hand if the
change in chemical
 potential is maintained for a long time, the
system explores other
 metastable states, the charge being allowed to
change in the
 process, thereby leading to a finite response. The
mean-field
 solution presumably overestimates the separation between
metastable
 states. Transitions can occur through activated processes
due to
 quantum or thermal fluctuations (so called creep). Although
treating
 such processes is difficult a possible modification of
(\ref{eq:compag}) is
\begin{equation} \label{eq:scaling}
\kappa(q;t,t_w) = \frac{\rho_0^2}c F\left[q L(t_w), h(t)/h(t_w)\right]
\end{equation}
with $F(0,y)=0$, $F(x,\infty)=F(\infty,y)=1$. Classical creep
arguments~\cite{Leon}
based on barriers growing as $L^{\theta}$ suggest $L(t)=L_c + h(t)$
with $h(t)=(T \ln t)^{2/\theta}$. Extensions incorporating quantum
effects, as in \cite{nattermann_temperature_luttinger}, are needed to
complete this picture.

In addition, the approach developed here allows one
to generalize the mode-coupling theory to
low-temperature glasses with quantum fluctuations and no quenched
disorder (see \cite{cugliandolo_dynamics_leshouches} for a discussion of the
classical limit).

We acknowledge interesting discussions with Z. Ovadyahu, A. K.
Savchenko, D. Reichman,
and A. Yacoby. This work was supported in part by the
Swiss National Science Foundation under MANEP and Division II, and
an ACI-France grant. LFC is a member of IUF.


\end{document}